\documentclass[aps,prc,twoside,twocolumn,fleqn,floatfix,showpacs,
superscriptaddress]{revtex4}
\usepackage{amssymb}
\usepackage{bm}
\usepackage{graphicx}

\def\cle{Laboratoire de Physique Corpusculaire, IN2P3/CNRS and 
Universit\'e Blaise Pascal, Clermont-Ferrand, France}
\def\buc{National Institute for Physics and Nuclear Engineering, Bucharest, 
Romania}
\def\bud{KFKI Research Institute for Particle and Nuclear Physics, Budapest, 
Hungary}
\def\gsi{Gesellschaft f\"ur Schwerionenforschung, Darmstadt, Germany }
\def\dre{IKH, Forschungszentrum Rossendorf, Dresden, Germany}
\def\hei{Physikalisches Institut der Universit\"at Heidelberg, Heidelberg, 
Germany }
\def\mos{Institute for Theoretical and Experimental Physics, Moscow, Russia }
\def\kur{Russian Research Center ``Kurchatov Institute", Moscow, Russia }
\def\kor{Korea University, Seoul, South Korea }
\def\str{Institut de Recherches Subatomiques, IN2P3-CNRS and 
Universit\'{e} Louis Pasteur, Strasbourg, France }
\def\pol{Institute of Experimental Physics, Warsaw University, Warsaw, 
Poland}
\def\zag{Rudjer Boskovic Institute, Zagreb, Croatia}
\def\sac{Service de Physique Th\'eorique, CEA-Saclay, Gif-sur-Yvette, 
France}

\begin{document}

\title[]{First analysis of anisotropic flow with Lee--Yang zeroes}

\author{N.~Bastid} \email{nicole.bastid@clermont.in2p3.fr} \affiliation{\cle} 
\author{A.~Andronic} \affiliation{\gsi}
\author{V. Barret} \affiliation{\cle}
\author{Z. Basrak} \affiliation{\zag}
\author{M.L. Benabderrahmane} \affiliation{\hei}
\author{R.\v{C}aplar} \affiliation{\zag}
\author{E. Cordier} \affiliation{\hei}
\author{P. Crochet} \affiliation{\cle} 
\author{P. Dupieux} \affiliation{\cle} 
\author{M. D\v{z}elalija} \affiliation{\zag}
\author{Z. Fodor} \affiliation{\bud}
\author{I. Ga\v{s}pari\'c} \affiliation{\zag}
\author{A. Gobbi} \affiliation{\gsi}
\author{Y. Grishkin} \affiliation{\mos}
\author{O.N. Hartmann} \affiliation{\gsi}
\author{N. Herrmann} \affiliation{\hei} 
\author{K.D. Hildenbrand} \affiliation{\gsi}
\author{B. Hong} \affiliation{\kor}
\author{J. Kecskemeti} \affiliation{\bud}
\author{Y.J. Kim} \affiliation{\kor} \affiliation{\gsi}
\author{M. Kirejczyk} \affiliation{\gsi} \affiliation{\pol}
\author{P. Koczon} \affiliation{\gsi}
\author{M. Korolija} \affiliation{\zag}
\author{R. Kotte} \affiliation{\dre}
\author{T. Kress} \affiliation{\gsi}
\author{A. Lebedev} \affiliation{\mos}
\author{Y. Leifels} \affiliation{\gsi}
\author{X. Lopez} \affiliation{\cle}
\author{A. Mangiarotti} \affiliation{\hei}
\author{V. Manko} \affiliation{\kur}
\author{M. Merschmeyer} \affiliation{\hei}
\author{D. Moisa} \affiliation{\buc}
\author{W. Neubert} \affiliation{\dre}
\author{D. Pelte} \affiliation{\hei}
\author{M. Petrovici} \affiliation{\buc}
\author{F. Rami} \affiliation{\str}
\author{W. Reisdorf}\affiliation{\gsi}
\author{A. Schuettauf} \affiliation{\gsi}
\author{Z. Seres} \affiliation{\bud}
\author{B. Sikora} \affiliation{\pol}
\author{K.S. Sim} \affiliation{\kor}
\author{V. Simion} \affiliation{\buc}
\author{K. Siwek-Wilczy\'nska} \affiliation{\pol}
\author{M.M. Smolarkiewicz} \affiliation{\pol}
\author{V. Smolyankin} \affiliation{\mos}
\author{I.J. Soliwoda} \affiliation{\pol}
\author{M.R. Stockmeier} \affiliation{\hei}
\author{G. Stoicea} \affiliation{\buc}
\author{Z. Tyminski} \affiliation{\gsi} \affiliation{\pol}
\author{K. Wi\'sniewski} \affiliation{\pol}
\author{D. Wohlfarth} \affiliation{\dre}
\author{Z. Xiao} \affiliation{\gsi}
\author{I. Yushmanov} \affiliation{\kur}
\author{A. Zhilin} \affiliation{\mos}

\collaboration{FOPI Collaboration}
\noaffiliation

\author{J.-Y.~Ollitrault} \affiliation{\sac} 
\author{N.~Borghini} \affiliation{\sac}

\date[]{Received April 1, 2005}

\begin{abstract}
We report on the first analysis of directed and elliptic flow with the 
new method of Lee--Yang zeroes. 
Experimental data are presented for Ru+Ru reactions 
at 1.69$A$~GeV measured with the FOPI detector at SIS/GSI. 
The results obtained with several methods, based on the event-plane 
reconstruction, on Lee--Yang zeroes, and on multi-particle 
cumulants (up to 5th order) applied for the first time 
at SIS energies, are compared. 
They show conclusive evidence that azimuthal correlations 
between nucleons and composite particles at this energy are largely dominated  
by anisotropic flow. 
\end{abstract}


\maketitle

The study of collective flow in relativistic heavy ion reactions is of great 
interest since it is expected to shed light on our knowledge about the 
properties of hot and dense nuclear matter and the underlying equation of 
state (EoS) \cite{dan02}. As pointed out early on, nuclear collective 
flow is also influenced by the momentum-dependent interactions and 
the in-medium nucleon-nucleon cross section \cite{dan93,li99}. 
Both effects play a crucial role in the determination of the EoS and 
cannot be neglected at intermediate energies. 
In this regard both directed and elliptic flow are a field of intense 
experimental and theoretical researches (see \cite{rei97} and 
references therein). 

Most flow analyses, based 
either on the reaction plane reconstruction 
(the so-called event-plane method) \cite{dan85}  
or on two-particle azimuthal correlations \cite{wan91} rely on 
the assumption that the only correlations are those stemming 
from the existence of the reaction plane. 
Other correlations (usually called non-flow), 
such as small-angle correlations 
due to final state interactions and quantum statistical effects \cite{kot04}, 
correlations due to resonance 
decays \cite{hon97} and mini-jet production \cite{kov02} are neglected. 
In recent years, several alternative techniques were introduced,
in which non-flow correlations can be unraveled. The 
cumulant method is based on a cumulant expansion of multi-particle 
(typically four particles) correlations \cite{bor01}, which  eliminates
most of non-flow correlations. 
It has been applied at ultra-relativistic energies, at RHIC and SPS 
for directed and elliptic flow studies and also for higher harmonic 
measurements \cite{adl02,alt03,ada04,tan04,phe04}. 
More recently, a new method based on an analogy with 
the Lee--Yang theory of phase transitions \cite{lee52}, 
where flow is extracted directly from the 
genuine correlation between a large number of particles, 
has been proposed \cite{zer1,zer11,zer2}. 
This method is expected to provide the cleanest 
separation between flow and non-flow effects. 

We present the first analysis of collective 
flow using the new method of Lee--Yang zeroes. 
The cumulant method is also applied, for the first time at 
SIS energies. A comparison with results obtained with 
the event-plane method is performed. 
We are thus able to check for the first time 
the validity of standard methods at SIS energies, 
by investigating possible contributions of correlations 
unrelated to the reaction plane which could introduce distortions on 
directed and elliptic flow results. 

The data set presented in this work concerns Ru+Ru reactions at 1.69$A$~GeV 
measured with the 
FOPI detector installed at the SIS accelerator facility of GSI-Darmstadt. 
FOPI is an azimuthally symmetric 
apparatus made of several sub-detectors which provide charge and 
mass determination over 
nearly the full $4\pi$ solid angle. The central part 
($ 33^\circ < \theta_{\rm lab} < 150^\circ$) is 
placed in a super-conducting solenoid and consists of a drift chamber (CDC) 
surrounded by a barrel of plastic scintillators. Particles measured in 
the CDC are identified by their mass using 
magnetic rigidity and energy loss. The forward part is composed of a wall 
of plastic scintillators 
($1.2^\circ < \theta_{\rm lab} < 30^\circ$) and an other drift chamber 
(Helitron) mounted inside the 
super-conducting solenoid. The plastic wall provides charge 
identification of the reaction 
products, combining time of flight and specific energy loss informations. 
For the present analysis, the forward wall and the CDC were used. 
More details on the configuration and performances of 
the different components of the FOPI apparatus can be 
found in \cite{fopi1}. 

The events are sorted out according to their degree of centrality 
by imposing conditions on the multiplicity of charged particles 
measured in the outer plastic wall 
($7^\circ < \theta_{\rm lab} < 30^\circ$) \cite{ala92}, named PMUL. 
The flow analysis presented here was carried out for 
about 2.9 million events belonging to the centrality 
class labelled PMUL4, which corresponds to a mean geometrical
impact parameter of 2.9 fm and to a geometrical impact parameter 
range from 1.6 fm to 3.9 fm, obtained assuming a sharp-cut-off 
approximation \cite{cav90}. 

We recall that directed flow ($v_1$) and elliptic flow ($v_2$) are 
quantified by Fourier coefficients of the azimuthal
distributions \cite{Voloshin:1994mz}, 
$ v_n = \langle \cos n(\varphi-\varphi_R)\rangle$, 
where $\varphi$ is the particle azimuthal angle and $\varphi_R$ is 
the azimuth of the reaction plane. 

In the conventional method, the reaction plane is estimated 
event by event according to the standard 
transverse momentum procedure devised in \cite{dan85}, which allows to 
construct the event-plane vector
\begin{equation}
\label{defQ}
{\bf Q}=\sum_\nu \omega_\nu {\bf u}_\nu. 
\end{equation}
The sum runs over all charged particles in the event, 
except pions identified in the CDC. 
${\bf u}_\nu$ is the unit vector parallel to the particle 
transverse momentum 
(i.e. ${\bf u}_\nu=(\cos\varphi_\nu,\sin\varphi_\nu)$, where 
$\varphi_\nu$ is the particle azimuth), and 
$\omega_\nu$ is a weight to improve the resolution, depending on the scaled 
center-of-mass (c.m.) rapidity $y^{(0)}= (y/y_p)_{\rm c.m.}$ (the 
subscript $p$ refers to the projectile):
$\omega_\nu=-1$ for $y^{(0)}<-0.3$, 
$\omega_\nu=+1$ for $y^{(0)}>0.3$, and
$\omega_\nu=0$ otherwise. 
The azimuth of ${\bf Q}$, denoted by $\Psi_R$, is 
an estimate of $\varphi_R$. 

The Fourier coefficients $v_n$ are calculated 
using the formula 
\begin{equation}
v_n\{EP\}\equiv\frac{\langle\cos n(\varphi-\Psi_R)\rangle}
{\langle\cos n\Delta\varphi_R\rangle}, 
\end{equation}
where $\{EP\}$ stands for ``event-plane''.  
In the numerator, the particle of 
interest is excluded from the sum in Eq.~(\ref{defQ}) 
to avoid autocorrelation effects. 
The resolution factor $1/\langle \cos n\Delta\varphi_R\rangle$ is an 
estimate of the error $\Delta\varphi_R=\Psi_R-\varphi_R$ 
on the determination of the reaction plane. 
This factor is calculated according to the procedure proposed in  
\cite{oll97}, and involves the correlation between randomly 
chosen sub-events \cite{dan85}. 
The numerical values for the PMUL4 centrality class are 
$1/\langle \cos \Delta\varphi_R\rangle=1.17$ for directed flow 
and 
$1/\langle \cos 2\Delta\varphi_R\rangle=1.69$ for elliptic flow, 
corresponding to a resolution parameter $\chi\approx 1.47$
\cite{oll97}. Several procedures have been developed in order 
to take into account correlations due to overall transverse 
momentum conservation, by using standard methods \cite{ogi89,bor03}. 
Here, to subtract these correlations the event-plane method has 
been improved by introducing a recoil correction, as proposed in \cite{ogi89}. 

Let us now recall the principle of the Lee--Yang zeroes procedure to 
analyze flow. 
A more complete description of the method 
can be found in \cite{zer1,zer11,zer2}. The method is based on the location 
of the zeroes, in the complex plane, of a generating function of 
azimuthal correlations, in close analogy with 
the theory of phase transitions of Lee and Yang \cite{lee52}. 
The first step of the procedure is to 
determine the ``integrated'' directed flow, defined as the 
average projection of ${\bf Q}$ on the (true) reaction plane 
\begin{equation}
\label{defintegrated}
V_1\equiv \langle Q_x \cos \varphi_R+Q_y \sin \varphi_R \rangle_{\rm events},
\end{equation}
where $Q_x$, $Q_y$ are the components of ${\bf Q}$, and 
the average is taken over events in a centrality class.
For this purpose, one introduces the following complex-valued generating 
function~\cite{zer2}:
\begin{equation}
\label{defG}
G^\theta(ir) = \bigg \langle \prod_{\nu} \lbrack 1 + i r \omega_\nu \cos 
(\varphi_\nu- \theta)\rbrack \bigg\rangle_{\rm events}
\end{equation}
where $r$ is a positive real variable, 
$\theta$ is an arbitrary reference angle, 
$\omega_\nu$ is the same weight as in Eq.~(\ref{defQ}), and 
$\varphi_\nu$ is the particle azimuthal angle. 

\begin{figure}[!hbt] 
\begin{center}
\includegraphics[width=8.5cm]{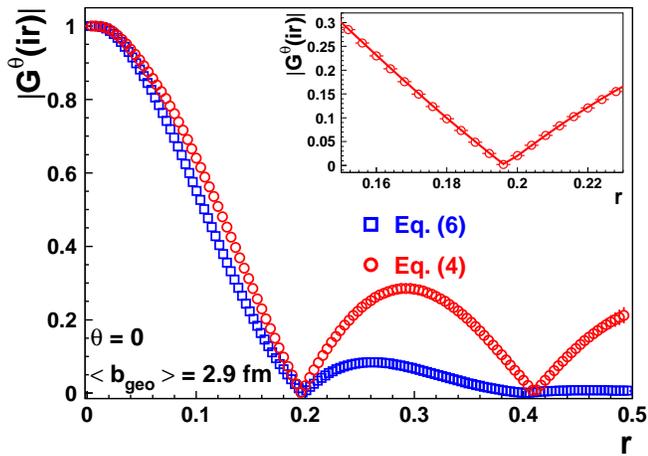}
\end{center}
\caption[]{$\left| G^\theta(ir)\right|$ {\it versus} $r$ 
for $\theta=0$. A zoom of $\left| G^\theta(ir)\right|$ (Eq.~(\ref{defG})) 
around its first minimum is shown in the insert and the solid line 
is just to guide the eye. See text for details.}
\label{fig-1} 
\end{figure}

Figure~\ref{fig-1} displays 
the amplitude of the generating function $|G^\theta(ir)|$ 
{\it versus} $r$ for  $\theta = 0$ (circles). 
It starts at a value of $1$ for $r$ = $0$ and quickly 
decreases as $r$ increases, which 
means that particles are strongly correlated:
for uncorrelated particles, indeed, $|G^\theta(ir)|$ is identically 
equal to unity within statistical fluctuations.
A sharp minimum of $|G^\theta(ir)|$ then occurs, which is in fact
compatible, within statistical fluctuations, with a zero of 
$G^\theta(ir)$ (see insert in Fig.~\ref{fig-1}). 
Following the general arguments presented in \cite{zer1,zer11,zer2}, 
this is a clear indication that correlations are due to 
collective flow. 
The position $r_0^\theta$ of the first minimum 
yields an estimate $V_1^\theta$ of the integrated flow $V_1$:
\begin{equation}
 V_1^\theta = {j_{01}\over r_0^\theta}, 
\end{equation}
where $j_{01} = 2.40483$ is the first root of the 
Bessel function $J_0(x)$.
Strictly speaking, like other flow analysis methods, the present one 
is only able to determine the absolute value of $V_1$. 
The sign is assumed to be positive at these energies. 

\begin{figure}[!hbt] 
\begin{center}
\includegraphics[width=8.6cm]{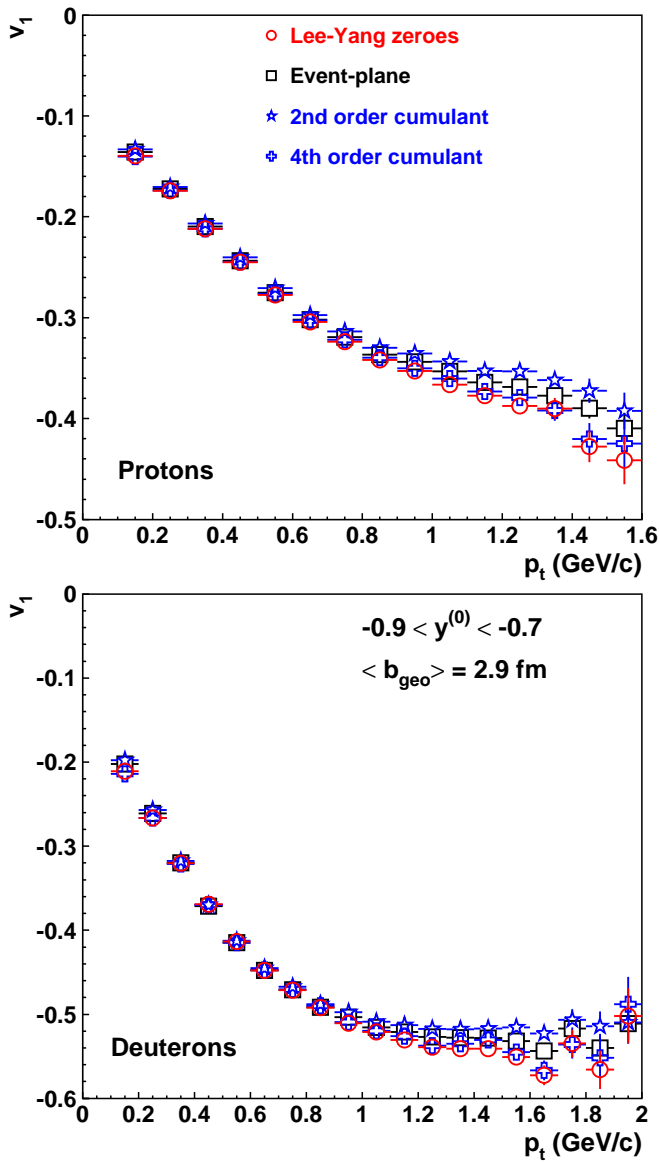}
\end{center}
\caption{$v_1$ {\it versus} transverse momentum for protons (upper panel) and 
deuterons (lower panel) measured in semi-central events and in a 
rapidity window in backward hemisphere. See text for details.}
\label{fig-2} 
\end{figure}

\begin{figure}[!hbt] 
\begin{center}
\includegraphics[width=8.6cm]{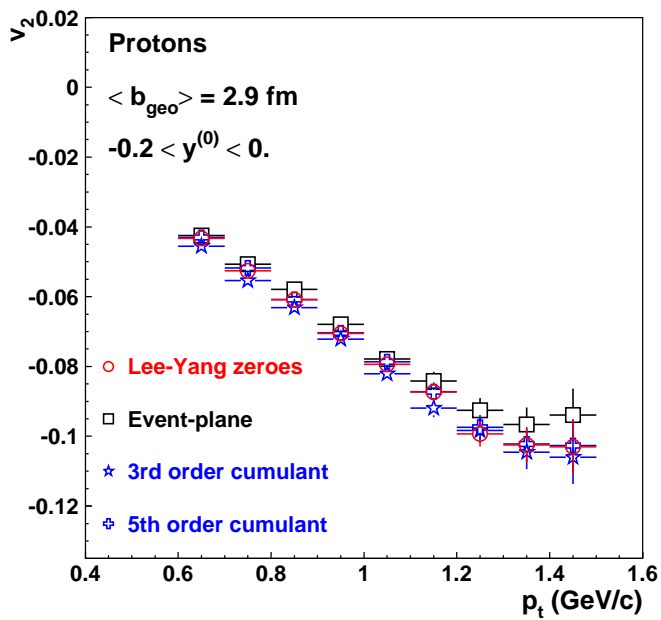}
\end{center}
\caption[]{$v_2$ {\it versus} transverse momentum 
for protons measured in semi-central events and around mid-rapidity. 
See text for details.
 }
\label{fig-3} 
\end{figure}

A potential limitation of the method comes from statistical errors,
which can be much larger than with the event-plane method. 
The reason why statistical errors depend on the method used is 
that the reaction plane is unknown, and that the $v_n$ are obtained 
through the indirect observation of a correlated emission
\cite{dan85}. The statistical errors depend on the observables
used to characterize this correlation. 
The important quantity here is the resolution parameter $\chi$, 
related to the well-known event-plane resolution \cite{oll97}. 
If $\chi>1$, which means that the reaction plane can be reconstructed
with reasonable accuracy, all methods yield statistical errors 
of the same order of magnitude as if the reaction plane was exactly
known while systematic errors from non-flow effects 
are expected to be much smaller with Lee--Yang zeroes. 
If $\chi<0.5$, on the other hand, statistical errors prevent the 
use of Lee--Yang zeroes. For the present analysis, we find $\chi\approx 1.45$, 
which definitely indicates that 
statistical errors are not a problem here (see Fig.~\ref{fig-2} and 
Fig.~\ref{fig-3}). 

An alternative form of the generating function, which can be 
used instead of Eq.~(\ref{defG}), is~\cite{zer1}:
\begin{equation}
\label{defGbis}
G^\theta(ir) = \left\langle \exp\left( i r\sum_{\nu} \omega_\nu \cos 
(\varphi_\nu- \theta) \right)\right\rangle_{\rm events}.
\end{equation}
This gives the squares in Fig.~\ref{fig-1}. 
There is no obvious relation between Eq.~(\ref{defG}) and
Eq.~(\ref{defGbis}). In particular, the latter differs from unity
for $r>0$ even if particles are uncorrelated because of 
``autocorrelation'' terms. 
But quite remarkably, the first minimum occurs at the same place 
with either form. 
This is a further indication that it is due to flow, 
as anticipated in Ref.~\cite{zer1} (see in particular Appendix A). 
The analysis of differential directed and elliptic flow presented 
below was carried out using Eq~(\ref{defG}). 
The results obtained with Eq.~(\ref{defGbis}) are very similar, 
thereby confirming that the method is insensitive to 
autocorrelations. 

Once the first minimum $r_0^\theta$ has been determined, 
the Fourier coefficients are estimated from the following equation:
\begin{equation}
v_n^\theta\propto{\rm Re} \bigg\langle\! \cos n(\varphi-\theta)
\prod_\nu  \phantom{ }' \lbrack 1 + i r_0^\theta \omega_\nu \cos 
(\varphi_\nu- \theta)\rbrack \bigg\rangle, 
\end{equation}
where $\varphi$ is the azimuthal angle of the analyzed particle, and 
the notation $\prod\phantom{ }'$ means that the particle of interest
is excluded from the product in order to avoid autocorrelations. 
The average is over a particle type in a given 
phase-space region,  in all events.
A proportionality constant ensures that the result is consistent
with the estimate of the integrated flow $V_1^\theta$. Its 
expression can be found in Ref.~\cite{zer2}. 

The procedure is repeated for several values of $\theta$ 
(typically, 5 equally spaced values 
from $0$ to $4\pi/5$), and the 
results are found to be independent of $\theta$ except for statistical 
fluctuations. This demonstrates that the results are not affected 
by detector azimuthal asymmetries. 
The final estimates shown below are averaged over $\theta$, 
which reduces the statistical errors by about a factor of 2. 

Before we come to the results, let us say a few words about the cumulant 
method \cite{bor01}, which has been already applied by several 
experiments \cite{adl02,alt03,ada04,tan04,phe04}. 
This method makes use of multi-particle 
correlations to estimate directed and elliptic flow. 
One can construct several independent estimates of 
$v_1$ and $v_2$, depending on how many particles are correlated: 
2 or 4 for $v_1$, 3 or 5 for $v_2$. The number of particles involved
is referred to as the order of the cumulant. 
Lowest-order estimates of $v_1$ and $v_2$ are not corrected for 
non-flow effects, and therefore are excepted to be similar to
estimates from the event-plane method without recoil correction.
The higher the order, the smaller the bias from 
non-flow correlations. Lee--Yang zeroes are essentially the limit
of cumulants when the order goes to infinity, and therefore 
minimize the bias from non-flow effects.  

The features of differential directed and elliptic flow at SIS energies 
have been discussed in several publications \cite{and01,and05}. 
Here, we focus on the comparison between 
the different procedures investigated in this work.

In the following figures only statistical errors are shown. 
Possible sources of systematic uncertainties have been studied using IQMD 
events \cite{aic91} passed through a complete GEANT simulation of the 
detector. We found that the 
full simulation underestimates $v_1$ of protons (deuterons) by 
about 6$\%$ (4$\%$), in relative value, in the 
phase space region under consideration and for data integrated over 
transverse momentum ($p_t$). These 
distortions are mainly due to a track-density effect which leads to a loss 
of particles in the directed flow direction. They are independent of the 
procedure. 

The differential directed flow calculated with the method of Lee--Yang 
zeroes (circles) is shown in 
Fig.~\ref{fig-2} for protons (upper panel) and deuterons (lower panel) 
in a rapidity window in the backward hemisphere. The values are compared 
to those obtained from the event-plane analysis (squares). 
Also shown are the second (stars) and fourth (crosses) 
particle cumulant values. 

A first look at Fig.~\ref{fig-2} shows that all methods give similar 
results. This proves that 
azimuthal correlations between nucleons and composite particles at SIS 
energies are dominated by anisotropic flow and that non-flow 
correlations, if any, are of smaller magnitude. Figure~\ref{fig-2} also 
shows that cumulants and Lee--Yang zeroes can be successfully used to 
analyze anisotropic flow 
at SIS energies. 

A detailed examination of the results in Fig.~\ref{fig-2} reveals however 
that there are small differences between the methods, beyond statistical 
errors at high $p_t$. 
First, there is a small difference between the event-plane method and
the second-order cumulant. This is due to the recoil correction 
for overall momentum conservation, which is applied in the event-plane 
method, but not in the cumulant method. We have checked that the 
event-plane method without recoil correction and the second-order 
cumulant give compatible results. 
It is important to emphasize that four-particle cumulants 
and event-plane results differ. 
The difference between second-order cumulant and 
fourth-order cumulant is also observed in analyses of elliptic 
flow at RHIC \cite{adl02}, where discrepancies are larger 
in relative value. 
There, it was suggested that the difference may be due 
to fluctuations of the flow within the sample of events, 
corresponding to variations in 
the impact parameter or in the initial conditions~\cite{adl02,mil03}. 
This effect leads to smaller flow estimates with
the four-particle cumulant than with the second-order cumulant, 
in absolute value, independently of $p_t$. 
The opposite behavior is evidenced in Fig.~\ref{fig-2}, from which 
one concludes that such fluctuations are not responsible for the 
observed differences. 
It seems therefore more likely that they are due to 
non-flow correlations. The fact that they increase with $p_t$ suggests
that they are mostly due to overall transverse momentum 
conservation~\cite{bor03a}.
The recoil correction which has been used in the 
event-plane method to correct 
for this effect relies on a non-relativistic formalism~\cite{ogi89} 
and that may explain the difference relative to the fourth-order cumulant. 
Moreover, it is worth to stress the fact 
that overall momentum conservation, 
which is a long-range effect involving all particles, effectively behaves 
as a short-range correlation \cite{zer1,bor03a}. As a 
consequence it is eliminated by using fourth-order cumulant. 
On the other hand, results from four-particle cumulants and 
Lee--Yang zeroes are perfectly compatible. This lends support to the
idea that both methods are able to extract reliably the genuine collective 
flow at SIS energies. 

Figure~\ref{fig-3} displays the proton differential elliptic flow, 
estimated from the same methods. 
The results now concern the mid-rapidity region 
($ -0.2 < y^{(0)} < 0.$) and the corresponding 
$p_t$ range of the CDC acceptance. 
Effects of non-flow correlations 
such as correlations due to momentum conservation are expected to be 
less pronounced on elliptic flow than on directed flow. 
Indeed, the differences between the methods are smaller, in absolute 
value, than for directed flow and can be considered as almost negligible 
within statistical error bars. 
However the general trend seems to be that the lowest-order 
cumulant (stars) gives a slightly larger signal than Lee--Yang zeroes 
(circles) and fifth-order cumulant (crosses), at all $p_t$. 
This small difference 
could be due to impact parameter fluctuations \cite{adl02} 
within the PMUL4 centrality bin which increase the 
estimates of $v_2$ from cumulants, 
in absolute value, independently of $p_t$. This bias is expected to be 
more pronounced for the third-order cumulant than for high-order cumulants. 
$v_2$ values from the event-plane 
method (squares) are not distorted by such fluctuations because of 
the high accuracy 
on the reaction plane determination. 
Similar trends have been also obtained for deuterons.

In summary, we have presented the first analysis of directed 
and elliptic flow in heavy-ion collisions using the method
of Lee--Yang zeroes. Results were obtained from the FOPI experiment 
at GSI. Such method is expected to provide the best
possible separation between correlations due to flow and 
other correlations. We were thus able to check explicitly that most azimuthal 
correlations between protons and composite particles
at SIS energies are due to their correlation with the 
reaction plane of the collision. 
There is no evidence for event-by-event fluctuations of 
directed flow. 
Non-flow effects are small; they are clearly seen only on 
directed flow at high $p_t$, and may be entirely ascribed to global 
transverse momentum conservation. 
They are eliminated using four-particle cumulants or the 
Lee--Yang zeroes procedure. 
Results were presented only for semi-central events, for sake 
of brevity. The analysis was also carried out 
for other centrality classes, covering an 
impact parameter range up to 7 fm, and led to similar 
conclusions. 

Such analysis is promising for studying pion flow.  
Since most pions originate from $\Delta$ decays, 
the resulting non-flow correlations with protons may 
contaminate the flow analysis if standard procedures 
are applied. The Lee--Yang zeroes method, which is insensitive
to non-flow effects, should be able to provide reliable results. \\

\begin{acknowledgments}
This work was partly supported by the agreement between GSI and IN2P3/CEA.
\end{acknowledgments}

\end{document}